# Power-Aware Digital Twin of Coherent Optical Receiver

Ambashri Purkayastha [1,2,3], Camille Delezoide [1], Vinod Bajaj [1], Fabien Boitier [1], Mounia Lourdiane [2], Cédric Ware [3], Patricia Layec [1]

[1] Nokia Bell Labs, France; [2] SAMOVAR, Télécom SudParis, Institut Polytechnique de Paris, France; [3] LTCI, Télécom Paris, Institut Polytechnique de Paris, France
Author e-mail address: ambashri.purkayastha@nokia.com

***Abstract:*** *We propose a digital twin of coherent receivers based on an extended physical model for predicting quality of transmission under receiver power variations. We experimentally validate it, demonstrating an accuracy improved by up to 1.5dB.*

## I. INTRODUCTION

Optical network digital twins (DTs) have been proposed for improving lightpaths' quality of transmission (QoT) predictions through monitoring data [1-3]. Twins further provide a simulation platform for sandboxing changes before deployment, such as power adjustments [4]. In typical coherent lightpaths, QoT prediction accuracy is mainly driven by the knowledge of the line. Thus, most proposed DT architectures focus on key line characteristics [1], such as amplifier's performances and fiber nonlinearities. Yet, coherent technology is progressively deployed in shorter lightpaths such as data-center interconnects where the QoT is chiefly limited by transceivers, especially if they are low-cost. This calls for higher accuracy when accounting for transceiver performances within lightpath's QoT prediction models.

Coherent transceivers are typically calibrated in "back-to-back" for each supported configuration, e.g. a modulation format and baudrate, to get the SNR-GOSNR curve [5]: the signal-to-noise ratio (SNR) before decision [6] - accounting for all noises - with respect to the generalized optical signal-to-noise ratio (GOSNR), only accounting for line noises. The cumulated noise contribution from both transmitter (Tx) and receiver (Rx) is deduced from the curve's high-OSNR asymptote. In the present mode of operation (PMO), this contribution is assumed constant throughout the lifetime of the network. However, it significantly depends on receiver's input power which may largely vary over time, as we observed on field data [7], further demonstrating that accounting for power variations doubled the accuracy of SNR predictions. Parallelly, in [8], Mano *et al* proposed a physics-based model accounting for photodetection noises to describe the impact of the receiver's input power ($P_{rx}$) on transceiver performance.

In this paper, we propose an extended model within a coherent Rx DT to estimate and predict the SNR upon receiver input power variations. In section II, we present the DT's architecture and receiver model. In section III, we first perform experiments to validate the model. Then, we evaluate the accuracy of DT-enabled SNR predictions in a field scenario.

## II. ARCHITECTURE FOR POWER-AWARE DIGITAL TWIN OF COHERENT RECEIVERS

As described in Fig.1, coherent receivers mix the received optical signal with the local oscillator (LO) output prior to photodetection. Then, the four electrical signals are amplified (TIAs), quantized (ADCs) and digitally processed (DSP) before decision. The Rx may be built with or without automated gain control (AGC), which maintains ADCs' input powers at an optimal level regarding quantization noise. This is either achieved with an extra optical amplifier (RX OA in Fig. 1), or electrically through variable-gain TIAs [9,10]. In both instances, compensating for a low $P_{rx}$ comes with added amplification noise. In contrast, without AGC, low $P_{rx}$ results in relatively higher quantization noise.

A lightpath's GOSNR accounts for all line noises, i.e. amplified spontaneous emission (ASE) from line amplifiers and nonlinear noises. Yet, the lightpath's QoT is more accurately described by its SNR after DSP [11], which writes as:

$$\mathbf{SNR^{-1} = \alpha GOSNR^{-(1+\varepsilon)} + SNR_{trx}^{-1} = \alpha GOSNR^{-(1+\varepsilon)} + SNR_{tx}^{-1} + SNR_{rx}^{-1}} \qquad (1)$$

where α converts the GOSNR to the receiver's effective bandwidth and ε accounts for multiplicative noise. In contrast, $SNR_{trx}$ accounts for both transmitter ($SNR_{tx}$) and receiver ($SNR_{rx}$) noise contributions. From (1), we understand that the SNR is mainly driven by the $SNR_{trx}$ when GOSNR is much higher, e.g. by 10 dB and more, which corresponds to lightpaths with short transmission distances, and/or equipped with low-cost transceivers.

The transceiver DT proposed in Fig.1 is meant as a subpart of a lightpath DT [7] designed to estimate/predict its QoT through the SNR. It is built to estimate its $SNR_{trx}$ resulting from both $SNR_{tx}$ and $SNR_{rx}$. In practice however, $SNR_{tx}$ can be assumed as fully determined by the transceiver's configuration. In contrast, $SNR_{rx}$ through $P_{rx}$ can significantly fluctuate over time for many reasons: rerouting, power adjustment, fiber cuts and damages, etc. We thus focus here on twinning the receiver through live power measurements while $SNR_{tx}$ is assumed known from pre-deployment calibrations and datasheet information, as depicted in Fig. 1. The lightpath DT passes a configuration to the transceiver's DT to obtain the corresponding $SNR_{trx}$ and finally deduce the lightpath's SNR through (1). In live mode, the transceiver's DT provides an estimation of the current $SNR_{trx}$, allowing for better separation of line and transceiver noises. In the virtual mode mainly designed for sandboxing, planned updates are passed to the DT for prediction of upcoming SNR variations, which may impact the quality of service. For both live and virtual modes, the essential DT block is the physics-based receiver model that decomposes the $SNR_{rx}$ as an inverse sum of individual contributions from the components within the receiver block.

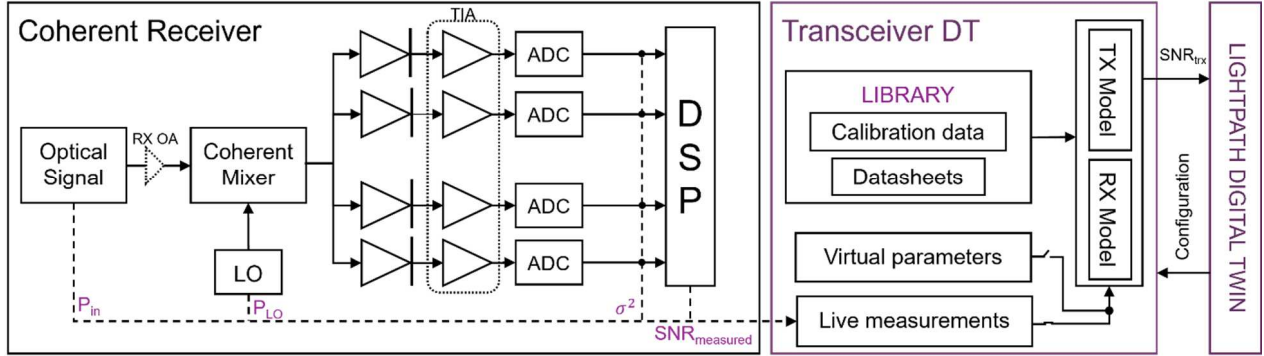

*Fig. 1. General architecture of optical coherent receiver connected to the transceiver digital twin.*

From [12], we get the receiver noise at the end of the TIA which includes the local oscillator noise term ($SNR_{LO}$) that combines the relative intensity noise and the laser phase noise, amplifier noise term from the TIAs ($SNR_{amp}$), and the photodetection noise ($SNR_{pd}$) that can be written as:

$$SNR_{pd}^{-1} = SNR_{sn}^{-1} + SNR_{DC}^{-1} + SNR_{th}^{-1} = \frac{2qB_{rx}RP_{LO} + 2qB_{rx}I_d + 4K_BTB_{rx}/R_L}{2R^2 P_{in}P_{LO}} = \frac{1}{P_{in}}C_{sn} + \frac{1}{P_{in}P_{LO}}(C_{th} + C_{DC}) \quad (2)$$

Here $SNR_{pd}$ includes a shot noise term ($SNR_{sn}$), a dark current term ($SNR_{DC}$), and a thermal noise term ($SNR_{th}$). The corresponding constants $C_{sn}$, $C_{dc}$ and $C_{th}$ depend on the elementary charge q, the dark current $I_d$, the Boltzmann constant $K_B$, and the resistor value $R_L$. $P_{in}$ and $P_{LO}$ are respectively the total optical input and LO powers. In practice, we ensure $P_{LO} \gg P_{in}$, to suppress thermal noise and ensure shot noise dominant condition. In [12], the receiver amplifier noise term contains only the thermal noise from the TIA, which is directly proportional to its noise figure and subsequently the gain. To this term, we add the contribution of the RX OA, which we assume adds similarly to the GOSNR. With AGC, $SNR_{amp}$ is proportional to $P_{in}$. Additionally, we include the quantization noise term (*SQNR*) from ADCs, that depends on the quantized signal variance $\sigma^2$ [13] which behaves as $\beta P_{in}$, where $\beta$ is the proportionality constant accounting for TIA gain and photodiode responsivity:

$$SQNR_{dB} = 10\log_{10}(\beta P_{in}) + SPS_{dB} - 10\log_{10}\Delta^2 + 10.79 \quad (3)$$

where $\Delta$ is the quantization step of the ADC, and SPS is the sampling rate. Let $SNR_{DSP}$ account for all DSP penalties. We can thus add the individual noise contributions from the different components, and write an expanded model for power sensitive receiver as:

$$SNR_{rx}^{-1} = SNR_{LO}^{-1} + SNR_{pd}^{-1} + SNR_{amp}^{-1} + SQNR^{-1} + SNR_{DSP}^{-1} \quad (4)$$

### III. EXPERIMENTAL MODEL VALIDATION AND APPLICATION OF THE DT FOR QOT PREDICTION

The experimental setup shown in Fig. 2a was configured for 32 GBaud DP-QPSK signals. The receiver integrates fixed gain TIAs and 8-bit ADCs. The noise constants $C_{sn}$, $C_{DC}$ and $C_{th}$ were calculated from the datasheet values of the 73 GHz balanced photodetectors. We monitor $P_{in1}$, $P_{in2}$, $P_{LO}$ and the total quantized signal power $\sigma^2$ after the ADCs. In Fig. 2b, we show the experimental SNR-GOSNR curve for $P_{in2}$ = 1.6 dBm and $P_{LO}$ = 16 dBm. $SNR_{trx}$, $\varepsilon$ and $\alpha$ are deduced from numerical fit, characterized by root-mean square error (RMSE) of 0.11 dB.

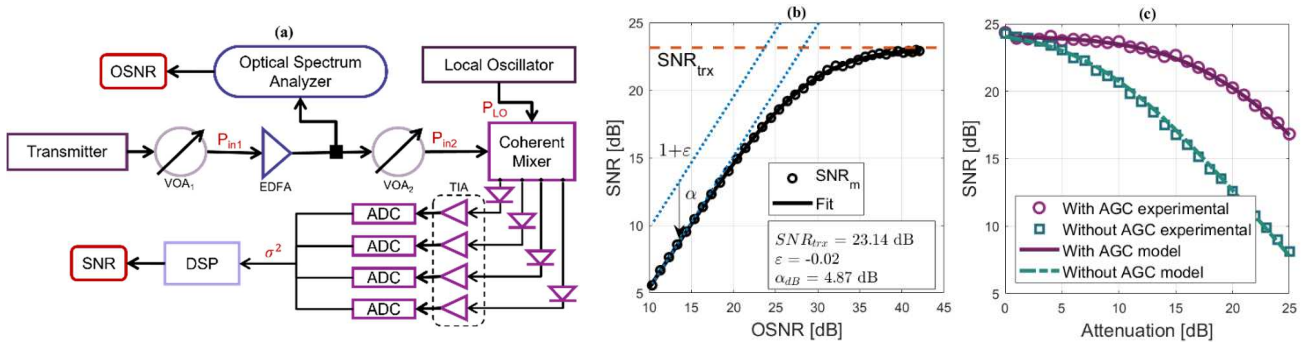

*Fig. 2. (a) Experimental Setup (b) Calibrated SNR-GOSNR curve (c) SNR evolution Vs power attenuation: with and without AGC.*

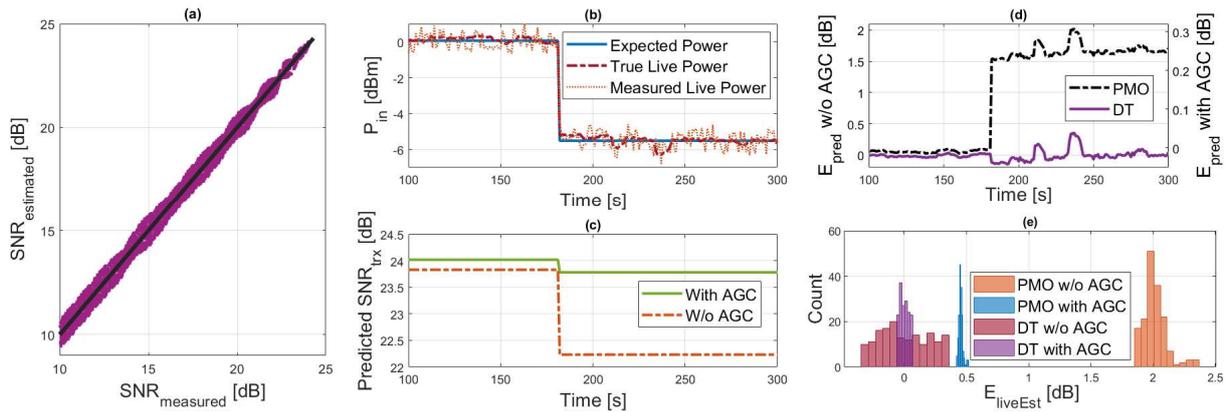

*Fig. 3. (a) Estimated SNR by the power aware DT for different measured SNR; (b) Expected, true and measured power evolution of lightpath under study (c) Evolution of predicted $SNR_{trx}$ for receiver with and without AGC; (d) SNR prediction error in virtual mode for PMO, and with power aware DT, at OSNR=36 dB for receiver with and without AGC; (e) Error distribution of estimated SNR in live mode for PMO and DT.*

To emulate a receiver with and without AGC we respectively vary $VOA_1$ and $VOA_2$, with RX OA emulated by an EDFA operated in constant power mode. In Fig. 2c, we present the evolution of SNR driven by $SNR_{amp}$ and SQNR. When AGC is applied, $SNR_{trx}$ reduces primarily due to added ASE noise. Without AGC, the ADC's response degrades from an actual 8-bit ADC at $P_{in2}$ = 1.6 dBm to the equivalent of a 5-bit ADC at $P_{in2}$ = -23 dBm, corresponding to an SQNR degradation of 18 dB. From the experimental data we numerically fit $\beta$, with an RMSE of 0.2 dB, demonstrating the validity of our receiver model that integrates quantization and amplifier noises in addition to photodetection noises.

The fitted parameters are added to the DT calibration library. The proposed DT is first tested in the live mode where monitoring data is used to estimate the current input-power dependent value of the $SNR_{trx}$. In Fig. 3a, we plot the estimated SNR with respect to the measured SNR for signals received at multiple input-power values. The std of the estimation error is 0.32 dB and the maximum error is 0.79 dB. The estimation error is lower in high SNR regime, since the model is less susceptible to measurement errors from the power monitoring devices. Then we simulate a case of dropped signal power during the lifetime of the lightpath as shown in Fig. 3b. The live power was collected from a lightpath from the field trial data in [14]. This respectively corresponds to $P_{in1}$ and $P_{in2}$ for the receiver with and without AGC. $SNR_{measured}$ is the SNR calculated using this live power, which we use as benchmark for comparing the predictions/estimations. Additionally, to emulate the live power input into the DT including measurement error by the power monitoring devices (e.g.: optical channel monitors), we added a uniformly distributed measurement noise between $\pm 0.8$ dB.

In Fig. 3c, we have the predicted evolution of the $SNR_{trx}$ with and without AGC. Without AGC, the $SNR_{trx}$ decreases by 1.6 dB when the power drops by 5.5 dB for the receiver. With AGC, $SNR_{trx}$ decreases response only by 0.3 dB. This $SNR_{trx}$ is then used to calculate the final SNR from (1) for OSNR = 36 dB, emulating short distance transmission performance, and the prediction is compared to the $SNR_{measured}$.

In Fig. 3d we plot the prediction error $E_{pred} = SNR_{predicted} - SNR_{measured}$ in dB, comparing the PMO—where $SNR_{trx}$ is assumed constant—with our proposed power-aware DT. With AGC, the default (PMO) prediction RMSE after the $P_{rx}$ drop is 0.25 dB. With the power-aware DT, the RMSE is reduced to 0.01 dB. Without AGC, the error is scaled due to higher $SNR_{trx}$ change, with the default RMSE of 1.68 dB, reduced to 0.1 dB with the power aware DT. Additionally, we have variation of prediction error, since it is blind to real power fluctuations in the network.

In Fig. 3e, we plot the distribution of the live estimation error $E_{liveEst} = SNR_{estimated} - SNR_{measured}$ in dB, when the DT has access to measurement of live power. In PMO, with AGC, the mean error $\mu = 0.45$ dB, with RMSE = 0.017 dB; without AGC, $\mu = 2.01$ dB and RMSE = 0.095. For the power-aware DT, with AGC the DT has $\mu = 0.001$ dB, and RMSE = 0.035 dB; and without AGC, $\mu = 0.0001$ dB, and RMSE = 0.19 dB. Increased standard deviation of the estimation is explained by measurement error from the monitoring systems.

IV. CONCLUSION

We have proposed a DT architecture aimed at QoT estimation/prediction. Our extended model integrates receiver amplifier noise, and the quantization noise within the transceiver noise to account for the impact of receiver power on the predicted SNR for transceiver noise limited systems. We then experimentally validate our model, obtaining an RMSE of 0.2 dB. Following this we built a digital twin of our experimental setup for two different receiver architectures—with and without AGC—on which we first predict the SNR performance under conditions of calibration and then predict the performance for a power drop of 5.5 dB, as observed in a field trial data. We show QoT prediction improvement of up to 1.5 dB. Additionally, we validate performance on live data, reducing QoT margins by up to 2 dB with the DT deployed when compared to the default mode of operation, and demonstrate that estimation is limited by the accuracy of monitoring systems.